\documentclass[journal]{IEEEtran}
\usepackage{cite}

  \usepackage{graphicx}
      \usepackage{epstopdf}
  \graphicspath{{../eps/}}
  \DeclareGraphicsExtensions{.eps}
\ifCLASSOPTIONcompsoc
  \usepackage[caption=false,font=normalsize,labelfont=sf,textfont=sf]{subfig}
\else
  \usepackage[caption=false,font=footnotesize]{subfig}
\fi

\usepackage{amsmath}
\usepackage{amssymb}
\usepackage{amsbsy}
\usepackage{amsthm}
\usepackage[linesnumbered,ruled]{algorithm2e}
\usepackage{url}

\usepackage[linesnumbered,ruled]{algorithm2e}
\usepackage{algcompatible}
\usepackage{booktabs} 
\usepackage[dvipsnames]{xcolor}

\newcommand{\tabincell}[2]{\begin{tabular}{@{}#1@{}}#2\end{tabular}}

\begin{document}
%
\title{Deep Attentive Features for \\ Prostate Segmentation in 3D Transrectal Ultrasound}
%
%
%

\author{Yi Wang*,
		Haoran Dou,
		Xiaowei Hu,
		Lei Zhu,
		Xin Yang,
		Ming Xu,
		Jing Qin,\\
		Pheng-Ann Heng,
		Tianfu Wang,
        and~Dong Ni
\thanks{This work was supported in part by the National Natural Science Foundation of China under Grants 61701312 and 61571304, in part by the Natural Science Foundation of SZU (No. 2018010), in part by the Shenzhen Peacock Plan (KQTD2016053112051497), and in part by a grant from the Research Grants Council of HKSAR (No. 14225616).
(Yi Wang and Haoran Dou contributed equally to this work.)
(Corresponding author: Yi Wang.)
}
\thanks{Y. Wang, H. Dou, T. Wang and D. Ni are with the National-Regional Key Technology Engineering Laboratory for Medical Ultrasound, Guangdong Key Laboratory for Biomedical Measurements and Ultrasound Imaging, School of Biomedical Engineering, Health Science Center, Shenzhen University, Shenzhen, China, and also with the Medical UltraSound Image Computing (MUSIC) Lab, Shenzhen, China (e-mail: onewang@szu.edu.cn).}
\thanks{X. Hu, X. Yang and P.A. Heng are with the Department of Computer Science and Engineering, The Chinese University of Hong Kong, Hong Kong, China.}
\thanks{L. Zhu and J. Qin are with the Centre for Smart Health, School of Nursing, The Hong Kong Polytechnic University, Hong Kong, China; L. Zhu is also with the Department of Computer Science and Engineering, The Chinese University of Hong Kong, Hong Kong, China.}
\thanks{M. Xu is with the Department of Medical Ultrasonics, the First Affiliated Hospital, Institute of Diagnostic and Interventional Ultrasound, Sun Yat-Sen University, Guangzhou, China.}
\thanks{Copyright (c) 2019 IEEE. Personal use of this material is permitted. However, permission to use this material for any other purposes must be obtained from the IEEE by sending a request to pubs-permissions@ieee.org.}
}

%
%

\markboth{Submit to IEEE TRANS. ON MEDICAL IMAGING,~Vol.~XX, No.~XX, XX~XX}
{Shell \MakeLowercase{\textit{et al.}}: Bare Demo of IEEEtran.cls for IEEE Journals}
%



\maketitle

\begin{abstract}
Automatic prostate segmentation in transrectal ultrasound (TRUS) images is of essential importance for image-guided prostate interventions and treatment planning.
However, developing such automatic solutions remains very challenging due to the missing/ambiguous boundary and inhomogeneous intensity distribution of the prostate in TRUS, as well as the large variability in prostate shapes.
This paper develops a novel 3D deep neural network equipped with attention modules for better prostate segmentation in TRUS by fully exploiting the complementary information encoded in different layers of the convolutional neural network (CNN).
Our attention module utilizes the attention mechanism to selectively leverage the multi-level features integrated from different layers to refine the features at each individual layer, suppressing the non-prostate noise at shallow layers of the CNN and increasing more prostate details into features at deep layers.
Experimental results on challenging 3D TRUS volumes show that our method attains satisfactory segmentation performance.
The proposed attention mechanism is a general strategy to aggregate multi-level deep features and has the potential to be used for other medical image segmentation tasks.
The code is publicly available at \url{https://github.com/wulalago/DAF3D}.
\end{abstract}

\begin{IEEEkeywords}
Attention mechanisms, deep features, feature pyramid network, 3D segmentation, transrectal ultrasound.
\end{IEEEkeywords}

%
\IEEEpeerreviewmaketitle

\section{Introduction}

\IEEEPARstart{P}{rostate} cancer is the most common noncutaneous cancer and the second leading cause of cancer-related deaths in men \cite{siegel2018cancer}.
Early detection and interventions is the crucial key to the cure of progressive prostate cancer \cite{pinto2011imaging}.
Transrectal ultrasound (TRUS) has long been a routine imaging modality for image-guided biopsy and therapy of prostate cancer \cite{hricak2007imaging}.
Accurate boundary delineation from TRUS images is of essential importance for the treatment planning \cite{wang2016towards}, biopsy needle placement \cite{yan2010discrete}, brachytherapy \cite{davis2012american}, cryotherapy \cite{bahn2002targeted}, and can help surface-based registration between TRUS and preoperative magnetic resonance (MR) images during image-guided interventions \cite{hu2012mr, wang2018online}.
Currently, prostate boundaries are routinely outlined manually in a set of transverse cross-sectional 2D TRUS slices,
then the shape and volume of the prostate can be derived from the boundaries for the subsequent treatment planning.
However, manual outlining is tedious, time-consuming and often irreproducible, even for experienced physicians.

\begin{figure}[t]
	\centering
	\includegraphics[width=0.95\linewidth]{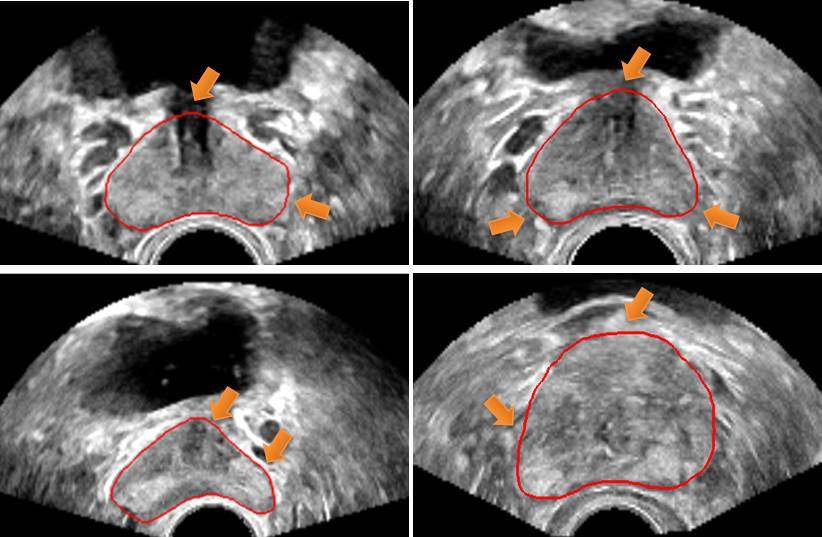}
	\caption{Example TRUS images. Red contour denotes the prostate boundary. There are large prostate shape variations, and the prostate tissues present inhomogeneous intensity distributions. Orange arrows indicate missing/ambiguous boundaries.}
	\label{fig:Challenges}
\end{figure}

Automatic prostate segmentation in TRUS images has become a considerable research area \cite{noble2006ultrasound, ghose2012survey}.
Nevertheless, even though there has been a number of methods in this area, accurate prostate segmentation in TRUS remains very challenging due to
(a) the ambiguous boundary caused by poor contrast between the prostate and surrounding tissues,
(b) missing boundary segments result from acoustic shadow and the presence of other structures (e.g. the urethra),
(c) inhomogeneous intensity distribution of the prostate tissue in TRUS images,
and (d) the large shape variations of different prostates~(see Fig.~\ref{fig:Challenges}).

\begin{figure*} [t]
	\centering
	\includegraphics[width=0.86\linewidth]{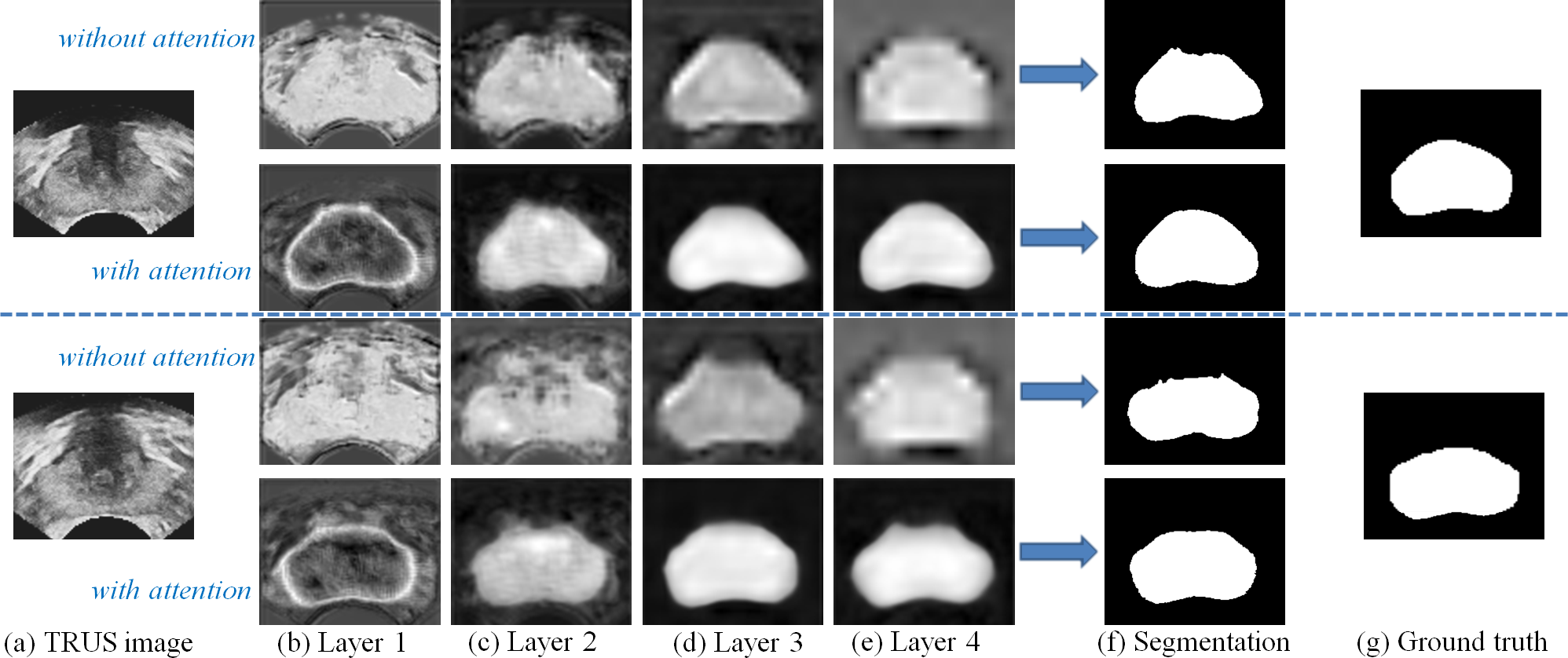}
	\caption{The visual comparisons of TRUS segmentations using conventional multi-level features (rows 1 and 3) and proposed attentive features (rows 2 and 4). (a) is the input TRUS images; (b)-(e) show the output feature maps from layer 1 (shallow layer) to layer 4 (deep layer) of the convolutional networks; (f) is the segmentation results predicted by corresponding features; (g) is the ground truths. We can observe that directly applying multi-level features without distinction for TRUS segmentation may suffer from poor localization of prostate boundaries. In contrast, our proposed attentive features are more powerful for the better representation of prostate characteristics.}
	\label{fig:mlf_daf}
\end{figure*}

\subsection{Relevant Work}
The problem of automatic prostate segmentation in TRUS images has been extensively exploited in the literature \cite{yan2010discrete, ladak2000prostate, pathak2000edge,  ghanei2001three, shen2003segmentation, wang2003semiautomatic, hu2003prostate, gong2004parametric, badiei2006prostate, tutar2006semiautomatic, zhan2006deformable, yann2011adaptively, ghose2013supervised, qiu2014prostate, santiago20152d, wu2015robust, li2016segmentation, yang20163d, zhu2017non, ma2017random}.
One main methodological stream utilizes shape statistics for the prostate segmentation.
Ladak \textit{et al.} \cite{ladak2000prostate} proposed a semi-automatic segmentation of 2D TRUS images based on shape-based initialization and the discrete dynamic contour (DDC) for the refinement.
Wang \textit{et al.} \cite{wang2003semiautomatic} further employed the DDC method to segment series of contiguous 2D slices from 3D TRUS data, thus obtaining 3D TRUS segmentation.
Pathak \textit{et al.} \cite{pathak2000edge} proposed a edge-guided boundary delineation algorithm with built-in a priori shape knowledge to detect the most probable edges describing the prostate.
Shen \textit{et al.} \cite{shen2003segmentation} presented a statistical shape model equipped with Gabor descriptors for prostate segmentation in ultrasound images.
Inspired by \cite{shen2003segmentation}, robust active shape model has been proposed to discard displacement outliers during model fitting procedure, and further applied to ultrasound segmentation \cite{santiago20152d}.
Tutar \textit{et al.} \cite{tutar2006semiautomatic} defined the prostate segmentation task as fitting the best surface to the underlying images under shape constraints learned from statistical analysis.
Yan \textit{et al.} \cite{yan2010discrete} developed a partial active shape model to address the missing boundary issue in ultrasound shadow area.
Yan \textit{et al.} \cite{yann2011adaptively} used both global population-based and patient-specific local shape statistics as shape constraint for the TRUS segmentation.
All aforementioned methods have incorporated prior shape information to provide robust segmentation against image noise and artifacts.
However, due to the large variability in prostate shapes, such methods may lose specificity, which are generally not sufficient to faithfully delineate boundaries in some cases \cite{ghose2012survey}.

In addition to shape statistics based methods, many other approaches resolve the prostate segmentation by formulating it as a foreground classification task.
Zhan \textit{et al.} \cite{zhan2006deformable} utilized a set of Gabor-support vector machines to analyse texture features for prostate segmentation.
Ghose \textit{et al.} \cite{ghose2013supervised} performed supervised soft classification with random forest to identify prostate.
Yang \textit{et al.} \cite{yang20163d} extracted patch-based features (e.g., Gabor wavelet, histogram of gradient, local binary pattern) and employed the trained kernel support vector machine to locate prostate tissues.
In general, all above methods used hand-crafted features for segmentations, which are ineffective to capture the high-level semantic knowledge, and thus tend to fail in generating high-quality segmentations when there are ambiguous/missing boundaries in TRUS images.

Recently, deep neural networks are demonstrated to be a very powerful tool to learn multi-level features for object segmentation \cite{ciresan2012deep, schmidhuber2015deep, long2015fully, ronneberger2015u, liskowski2016segmenting, havaei2017brain, chen2018deeplab}.
Guo \textit{et al.} \cite{7353170} presented a deep network for the segmentation of prostate in MR images.
Motivated by \cite{7353170}, Ghavami \textit{et al.} \cite{ghavami2018automatic, ghavami2018integration} employed convolutional neural networks (CNNs) built upon U-net architecture \cite{ronneberger2015u} for automatic prostate segmentation in 2D TRUS slices.
To tackle the missing boundary issue in TRUS images, Yang \textit{et al.} \cite{yang2017fine} proposed to learn the shape prior with the biologically plausible recurrent neural networks (RNNs) and bridged boundary incompleteness.
Karimi \textit{et al.} \cite{karimi2018accurate} employed an ensemble of multiple CNN models and a statistical shape model to segment TRUS images for prostate brachytherapy.
Anas \textit{et al.} \cite{anas2017clinical} employed a deep residual neural net with an exponential weight map to delineate the 2D TRUS images for low-dose prostate brachytherapy treatment.
Anas \textit{et al.} \cite{anas2018deep} further developed an RNN-based architecture with gated recurrent unit as the core of the recurrent connection to segment prostate in freehand ultrasound guided biopsy.

\begin{figure*} [t]
	\centering
	\includegraphics[width=0.88\linewidth]{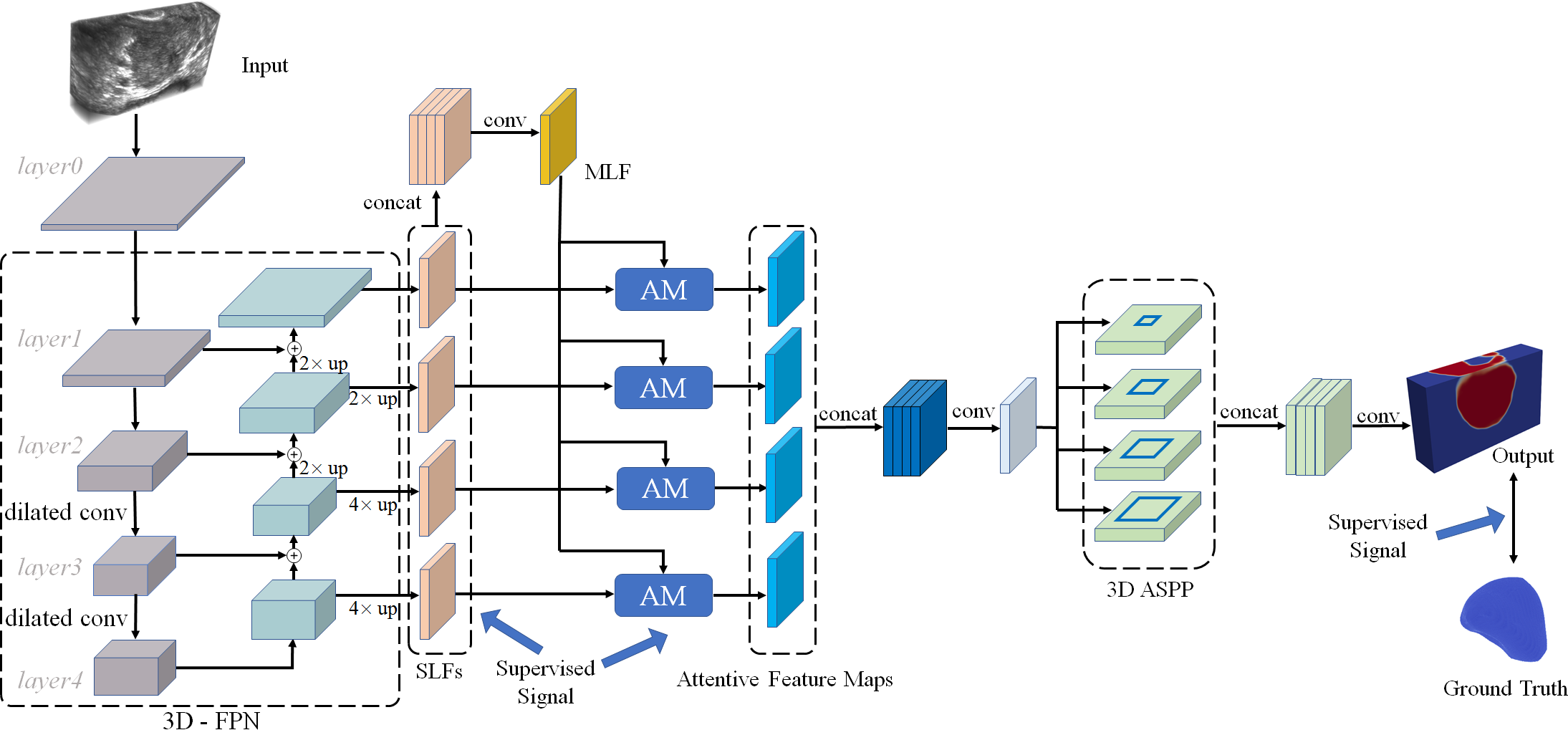}
	\caption{The schematic illustration of our prostate segmentation network equipped with attention modules. FPN: feature pyramid network; SLF: single-layer features; MLF: multi-layer features; AM: attention module; ASPP: atrous spatial pyramid pooling.}
	\label{fig:arc}
\end{figure*}

Compared to traditional machine learning methods with hand-crafted features, one of the main advantages of deep neural networks is to generate multi-level features consisting of abundant semantic and fine information.
However, directly applying multi-level convolutional features without distinction for TRUS segmentation may suffer from poor localization of prostate boundaries, due to the distraction from redundant features (see the 1st and 3rd rows of Fig.~\ref{fig:mlf_daf}).
Because the integrated multi-level features tend to include non-prostate regions (due to low-level details from shallow layers) or lose details of prostate boundaries (due to high-level semantics from deep layers) when generating segmentation results.
Our preliminary study on 2D TRUS images \cite{wang2018deep} has demonstrated that it is essential to leverage the complementary advantages of features at multiple levels and to learn more discriminative features targeting for accurate and robust segmentation.
However, the work \cite{wang2018deep} only realizes 2D segmentation which could be very limiting for its application.

In this study, we develop a novel 3D feature pyramid network equipped with attention modules to generate deep attentive features (DAF) for better prostate segmentation in 3D TRUS volumes.
The DAF is generated at each individual layer by learning the complementary information of the low-level detail and high-level semantics in multi-layer features (MLF), thus is more powerful for the better representation of prostate characteristics (see the 2nd and 4th rows of Fig.~\ref{fig:mlf_daf}).
Experiments on 3D TRUS volumes demonstrate that our segmentation using deep attentive features achieves satisfactory performance.

\subsection{Contributions}
The main contributions of our work are twofold.

\begin{enumerate}[]
	\item We propose to fully exploit the useful complementary information encoded in the multi-level features to refine the features at each individual layer.
	Specifically, we achieve this by developing an attention module, which can automatically learn a set of weights to indicate the importance of the features in MLF for each individual layer.

	\item We develop a 3D attention guided network with a novel scheme for TRUS prostate segmentation by harnessing the spatial contexts across deep and shallow layers.
	To the best of our knowledge, we are the first to utilize attention mechanisms to refine multi-layer features for the better 3D TRUS segmentation.
	In addition, the proposed attention mechanism is a general strategy to aggregate multi-level features and has the potential to be used in other segmentation applications.
	
\end{enumerate}

The remainder of this paper is organized as follow.
Section~\ref{sec:Method} presents the details of the attention guided network which generates attentive features by effectively leveraging the complementary information encoded in multi-level features.
Section~\ref{sec:Results} presents the experimental results of the proposed method for the application of 3D TRUS segmentation.
Section~\ref{sec:Discussions} elaborates the discussion of the proposed attention guided network, and the conclusion of this study is given in Section~\ref{sec:Conclusion}.

\section{Deep Attentive Features For 3D Segmentation}
\label{sec:Method}

Segmenting prostate from TRUS images is a challenging task especially due to the ambiguous/missing boundary and inhomogeneous intensity distribution of the prostate in TRUS.
Directly using low-level or high-level features, or even their combinations to conduct prostate segmentation may often fail to get satisfactory results.
Therefore, leveraging various factors such as multi-scale contextual information, region semantics and boundary details to learn more discriminative prostate features is essential for accurate and robust prostate segmentation.

To address above issues, we present deep attentive features for the better representation of prostate.
The following subsections present the details of the proposed scheme and elaborate the novel attention module.

\subsection{Network Architecture}
\label{sec:method_overview}

Fig.~\ref{fig:arc} illustrates the proposed prostate segmentation network with deep attentive features.
Our network takes the TRUS images as the input and outputs the segmentation result in an end-to-end manner.
It first produces a set of feature maps with different resolutions.
The feature maps at shallow layers have high resolutions but with fruitful detail information while the feature maps at deep layers have low resolutions but with high-level semantic information.
We implement the 3D ResNeXt \cite{xie2017aggregated} as the feature extraction layers (the gray parts in the left of Fig.~\ref{fig:arc}).
Specifically, to alleviate the issue of large scale variability of prostate shapes in different TRUS slices (e.g., mid-gland slices show much larger prostate region than base/apex slices do), we employ dilated convolution \cite{yu2015multi} in backbone ResNeXt to systematically aggregate multi-scale contextual information.
We use $3\times3\times3$ dilated convolutions with rate of $2$ to substitute the conventional $3\times3\times3$ convolutions in \textit{layer3} and \textit{layer4} to increase the receptive field without loss of resolution.
In addition, considering that the TRUS data is a ``thin'' volume (slice number ($L$) is relatively smaller than slice width ($W$)/height ($H$)), we set downsampling of \textit{layer0} by stride $(2,2,2)$, and set \textit{layer1}, \textit{layer2} by stride $(2,2,1)$ to retain useful information in different slices.

\begin{figure} [t]
	\centering
	\includegraphics[width=0.88\linewidth]{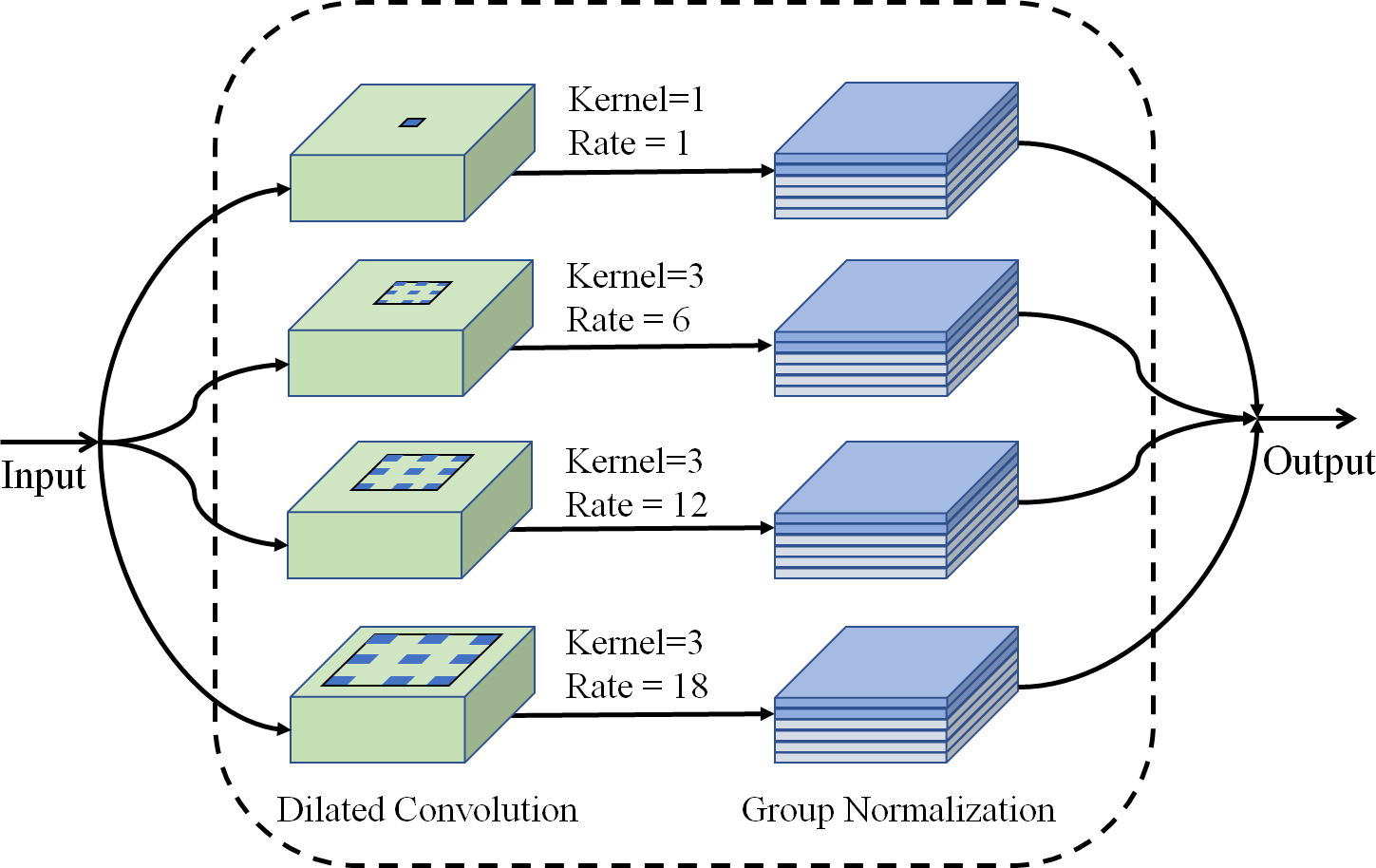}
	\caption{The schematic illustration of the atrous spatial pyramid pooling (ASPP) with dilated convolution and group normalization (GN).}
	\label{fig:aspp}
\end{figure}

To naturally leverage the feature hierarchy computed by convolutional network, we further utilize feature pyramid network (FPN) architecture \cite{lin2017feature} to combine multi-level features via a top-down pathway and lateral connections (see Fig.~\ref{fig:arc}, 3D-FPN).
The top-down pathway upsamples spatially coarser, but semantically stronger feature maps from higher pyramid levels.
These feature maps are then merged with correspondingly same-sized bottom-up maps via lateral connections.
Each lateral connection merges feature maps by element-wise addition.
The enhanced feature maps at each layer are obtained by using the deeply supervised mechanism \cite{xie2015holistically} that imposes the supervision signals to multiple layers.
The deeply supervised mechanism can reinforce the propagation of gradients flows within the 3D network and hence help to learn more representative features \cite{dou20173d}.
Note that the feature maps at \textit{layer0} are ignored in the pyramid due to the memory limitation.

After obtaining the enhanced feature maps with different levels of information via FPN, we enlarge these feature maps with different resolutions to the same size of \textit{layer1}'s feature map by trilinear interpolation.
The enlarged feature maps at each individual layer are denoted as ``single-layer features (SLF)'', and the multiple SLFs are combined together, followed by convolution operations, to generate the ``multi-layer features (MLF)''.
Although the MLF encodes the low-level detail information as well as the high-level semantic information of the prostate, it also inevitably incorporates noise from the shallow layers and losses some subtle parts of the prostate due to the coarse features at deep layers.

In order to refine the features of the prostate ultrasound image, we present an attention module to generate deep attentive features at each layer in the principle of the attention mechanism.
The proposed attention module leverages the MLF and the SLF as the inputs and produces the refined attentive feature maps; please refer to Section~\ref{sec:DAF} for the details of our attention module.

Then, instead of directly averaging the obtained multi-scale attentive feature maps for the prediction of the prostate region, we employ a 3D atrous spatial pyramid pooling (ASPP) \cite{chen2017rethinking} module to resample attentive features at different scales for more accurate prostate representation.
As shown in Fig.~\ref{fig:arc}, the multiple attentive feature maps generated by attention modules are combined together, followed by convolution operations, to form an attentive feature map.
Four parallel convolutions with different atrous rates are then applied on top of this attentive feature map to capture multi-scale information.
Specifically, the schematic illustration of our 3D ASPP with dilated convolution and group normalization (GN) \cite{Wu_2018_ECCV} is shown in Fig.~\ref{fig:aspp}.
Our 3D ASPP consists of
(a) one $1\times1\times1$ convolution and three $3\times3\times3$ dilated convolutions with rates of ($6, 12, 18$),
and (b) group normalization right after the convolutions.
We choose GN instead of batch normalization is due to GN's accuracy is considerably stable in a wide range of batch sizes \cite{Wu_2018_ECCV}, which will be more suitable for 3D data computation.
Our GN is along the channel direction and the number of groups is $32$.

Finally, we combine multi-scale attentive features together, and get the prostate segmentation result by using the deeply supervised mechanism \cite{xie2015holistically}.

\begin{figure} [t]
	\centering
	\includegraphics[width=0.55\linewidth]{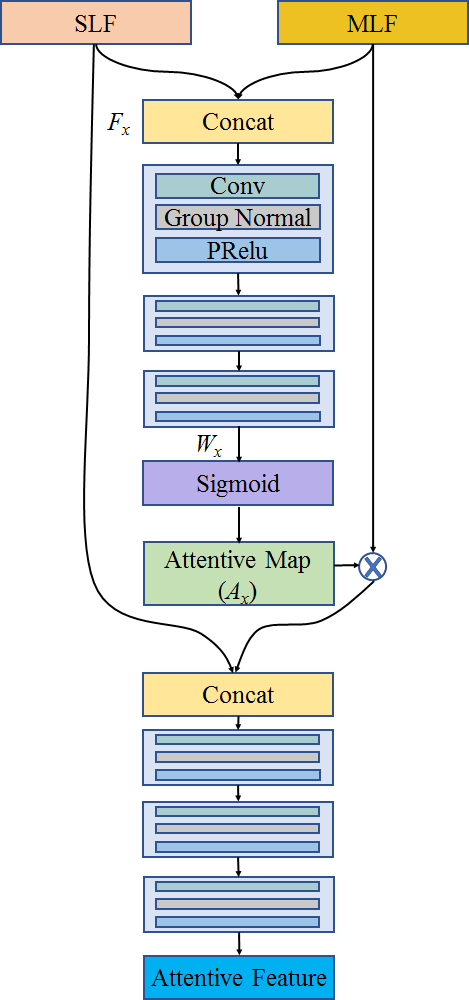}
	\caption{The schematic illustration of the proposed attention module.}
	\label{fig:daf}
\end{figure}

\subsection{Deep Attentive Features}
\label{sec:DAF}

As presented in Section~\ref{sec:method_overview}, the feature maps at shallow layers contain the detail information of prostate but also include non-prostate regions, while the feature maps at deep layers are able to capture the highly semantic information to indicate the location of the prostate but may lose the fine details of the prostate's boundaries.
In order to refine the features at each layer, here we present a deep attentive module (see Fig.~\ref{fig:daf}) to generate the refined attentive features by utilizing the proposed attention mechanism.

Attention model is widely used for various tasks, including image segmentation. Several attention mechanisms, e.g., channel-wise attention \cite{hu2018squeeze} and pixel-wise attention \cite{zhao2018psanet}, have been proposed to boost the network’s representational power.
In this study, we explore layer-wise attention mechanism to selectively leverage the complementary features across all scales to refine the features of individual layers.

Specifically, as shown in Fig.~\ref{fig:daf}, we feed the MLF and SLF at each layer into the proposed attention module and generate refined SLF through the following three steps.
The first step is to generate an attentive map at each layer, which indicates the importance of the features in MLF for each specific individual layer.
Given the single-layer feature maps at each layer, we concatenate them with the multi-layer feature maps as $F_x$, and then produce the unnormalized attention weights $W_{x}$ (see Fig.~\ref{fig:daf}):
\begin{equation}
\label{attention}
W_{x}=f_{a}(F_x;\theta),
\end{equation}
where $\theta$ represents the parameters learned by $f_a$ which contains three convolutional layers.
The first two convolutional layers use $3\times3\times3$ kernels, and the last convolutional layer applies $1\times1\times1$ kernels.
It is worth noting that in our implementation, each convolutional layer consists of one convolution, one group normalization, and one parametric rectified linear unit (PRelu) \cite{he2015delving}.
These convolutional operations are employed to choose the useful multi-level information with respect to the features of each individual layer.
After that, our attention module computes the attentive map $A_x$ by normalizing $W_{x}$ with a Sigmoid function.

In the second step, we multiply the attentive map $A_x$ with the MLF in a element-by-element manner to weight the features in MLF for each SLF.
Third, the weighted MLF is merged with corresponding features of each SLF by applying two $3\times3\times3$ and one $1\times1\times1$ convolutional layers, which is capable of automatically refining layer-wise SLF and producing the final attentive features for the given layer (see Fig.~\ref{fig:daf}).

In general, our attention mechanism leverages the MLF as a fruitful feature pool to refine the features of each SLF.
Specifically, as the SLF at shallow layers is responsible for discovering detailed information but lack of semantic information of prostate, the MLF can guide them gradually suppress details that are not located in the semantic saliency regions while capturing more details in semantic saliency regions.
Meanwhile, as SLF at deep layers are responsible for capturing cues of the whole prostate and may lack of detailed boundary features, the MLF can enhance their boundary details.
By refining the features at each layer using the proposed attention mechanism, our network can learn to select more discriminative features for accurate and robust TRUS segmentation.

\begin{figure} [t]
	\centering
	\includegraphics[width=0.85\linewidth]{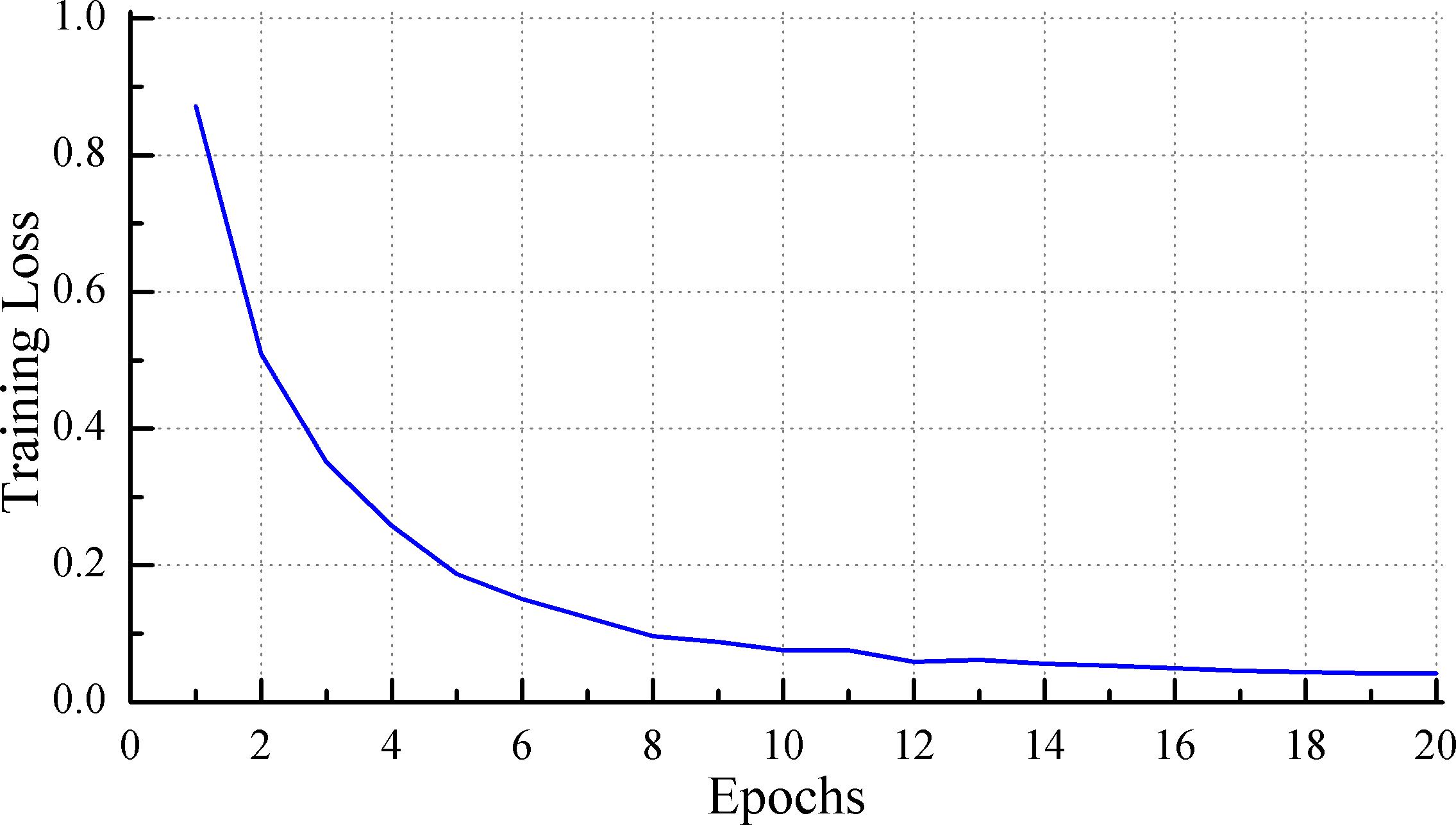}
	\caption{The learning curve of our attention guided network.}
	\label{fig:TrainLoss}
\end{figure}

\subsection{Implementation Details}
Our proposed framework was implemented on PyTorch and used the 3D ResNeXt \cite{xie2017aggregated} as the backbone network.

\begin{figure*} [t]
	\centering
	\includegraphics[width=0.9\linewidth]{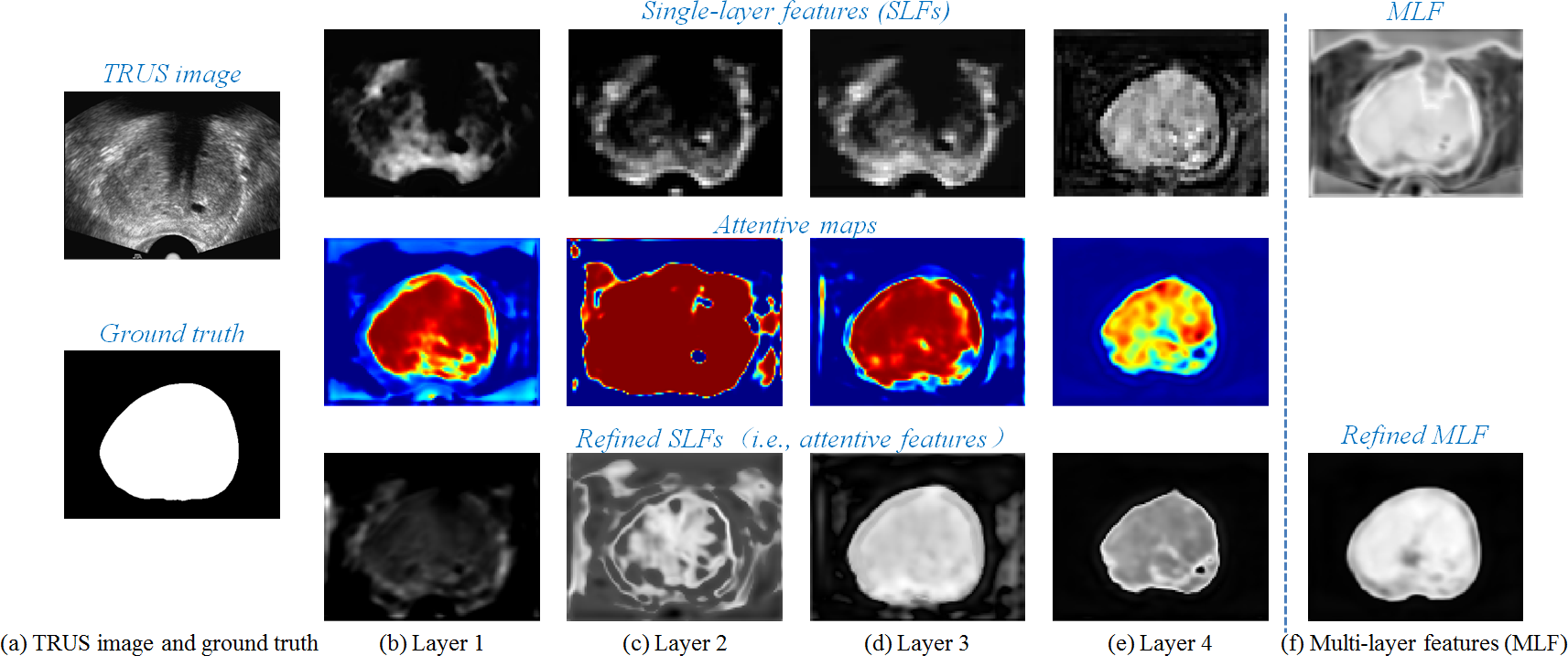}
	\caption{One example to illustrate the effectiveness of the proposed attention module for the feature refinement. (a) is the input TRUS image and its ground truth; (b)-(e) show the features from layer 1 (shallow layer) to layer 4 (deep layer); rows 1-3 show single-layer features (SLFs), corresponding attentive maps and attention-refined SLFs, respectively; (f) is the multi-layer features (MLF) and the attention-refined MLF. We can observe that our proposed attention module provides a feasible solution to effectively incorporate details at low levels and semantics at high levels for better feature representation.}
	\label{fig:attim}
\end{figure*}

\paragraph {\textbf{Loss Function}}
During the training process, Dice loss $\mathcal{L}_{dice}$ and binary cross-entropy loss $\mathcal{L}_{bce}$ are used for each output of this network:
\begin{equation}
\label{dice}
\mathcal{L}_{dice} = 1- \frac{2 \sum_{i=1}^{N} p_i g_i }{ \sum_{i=1}^{N} {p_i}^2 + \sum_{i=1}^{N} {g_i}^2 },
\end{equation}
\begin{equation}
\label{bce}
\mathcal{L}_{bce} = \sum_{i=1}^{N} g_i \log p_i + \sum_{i=1}^{N} (1-g_i) \log (1-p_i),
\end{equation}
where $N$ is the voxel number of the input TRUS volume;
$p_i \in [0.0, 1.0]$ represents the voxel value of the predicted probabilities;
$g_i \in \{0, 1\}$ is the voxel value of the binary ground truth volume.
The binary cross-entropy loss $\mathcal{L}_{bce}$ is a conventional loss in segmentation task.
It is preferred in preserving boundary details but may cause over-/under-segmentation due to class-imbalance issue.
In order to alleviate this problem, we combine the Dice loss $\mathcal{L}_{dice}$ with the $\mathcal{L}_{bce}$.
The Dice loss emphasizes global shape similarity to generate compact segmentation and its differentiability has been illustrated in \cite{milletari2016v}.
The combined loss is helpful to consider both local detail and global shape similarity.
We define each supervised signal $\mathcal{L}_{signal}$ as the summation of $\mathcal{L}_{dice}$ and $\mathcal{L}_{bce}$:
\begin{equation}
\label{signal}
\mathcal{L}_{signal} = \mathcal{L}_{dice} + \mathcal{L}_{bce}.
\end{equation}
Therefore the total loss $\mathcal{L}_{total}$ is defined as the summation of loss on all supervised signals:
\begin{equation}  \label{eq5}
\mathcal{L}_{total} = \sum_{i=1}^{n}w^i \mathcal{L}_{signal}^i + \sum_{j=1}^{n}w^j \mathcal{L}_{signal}^j + w^f \mathcal{L}_{signal}^f,
\end{equation}
where $w^i$ and $\mathcal{L}_{signal}^i$ represent the weight and loss of $i$-th layer;
while $w^j$ and $\mathcal{L}_{signal}^j$ represent the weight and loss of $j$-th layer after refining features using our attention modules;
$n$ is the number of layers of our network;
$w^f$ and $\mathcal{L}_{signal}^f$ are the weight and loss for the output layer.
We empirically set the weights ($w^{i=1,2,3,4}$, $w^{j=1,2,3,4}$ and $w^f$) as ($0.4, 0.5, 0.7, 0.8, 0.4, 0.5, 0.7, 0.8, 1$).

\paragraph{\textbf{Training Process}}
Our framework is trained end-to-end.
We adopt Adam \cite{Kingma2014Adam} with the initial learning rate of $0.001$, a mini-batch size of $1$ on a single TITAN Xp GPU, to train the whole framework.
Fig.~\ref{fig:TrainLoss} shows the learning curve of the proposed framework.
It can be observed that the training converges after $14$ epochs.
Training the whole framework by $20$ epochs takes about $54$ hours on our experimental data.

The code is publicly available at \url{https://github.com/wulalago/DAF3D}.

\section{Experiments and Results}
\label{sec:Results}

\subsection{Materials}
Experiments were carried on TRUS volumes obtained from forty patients at the First Affiliate Hospital of Sun Yat-Sen University, Guangzhou, Guangdong, China.
The study protocol was reviewed and approved by the Ethics Committee of our institutional review board and informed consent was obtained from all patients.

We acquired one TRUS volume from each patient.
All TRUS data were obtained using Mindray DC-8 ultrasound system (Shenzhen, China) with an integrated 3D TRUS probe.
These data were then reconstructed into TRUS volumes.
The 3D TRUS volume contains $170\times132\times80$ voxels with a voxel size of $0.5\times0.5\times0.5$ mm$^3$.
To insure the ground-truth segmentation as correct as possible, two experienced urological clinicians with extensive experience in interpreting the prostate TRUS images have been involved for annotations.
It took two weeks for one clinician to delineate all boundaries using a custom interface developed via C++.
This clinician delineated each slice by considering the 3D information of its neighboring slices.
Then all the manually delineated boundaries were further refined/confirmed by another clinician for the correctness assurance.
We adopted data augmentation (i.e., rotation and flipping) for training.

\subsection{Experimental Methods}
To demonstrate the advantages of the proposed method on TRUS segmentation, we compared our attention guided network with other three state-of-the-art segmentation networks: 3D Fully Convolutional Network (FCN) \cite{long2015fully}, 3D U-Net \footnote{The work \cite{ghavami2018automatic} adopted a 2D U-Net architecture \cite{ronneberger2015u} as backbone network. Here we extend \cite{ghavami2018automatic} to 3D architecture for a fair comparison.} \cite{ghavami2018automatic}, and Boundary Completion Recurrent Neural Network (BCRNN) \cite{yang2017fine}.
It is worth noting that the work \cite{yang2017fine} and \cite{ghavami2018automatic} have been proposed specializing in TRUS segmentation.
For a fair comparison, we re-trained all the three compared models using the public implementations and adjusted training parameters to obtain best segmentation results.

In addition to the aforementioned compared methods, we also performed ablation analysis to directly show numerical gains of the attention module design.
We discarded the attention modules in our framework, and directly sent the MLF (the yellow layer in Fig.~\ref{fig:arc}) to go through the ASPP module for the final prediction.
We denote this model as 3D customized FPN (cFPN).
Four-fold cross-validation was conducted to evaluate the segmentation performance of different models \footnote{To ensure fair comparison, same hyper-parameter tuning was conducted for each network in the cross-validation. More specifically, we sampled over a range over hyper-parameters and trained each network. Each network's performance shown in this paper was hyper-parameters that produced on average the best performance in all four folds.
}.

\begin{table*}[t]
	\caption{Metric results of different methods (Mean$\pm$SD, best results are highlighted in bold)}
	\label{table:state-of-the-art}
	\begin{center}
		\begin{tabular}{l|c|c|c|c|c}
			\toprule[1.5pt]
			Metric & 3D FCN~\cite{long2015fully} & 3D U-Net~\cite{ghavami2018automatic} & BCRNN~\cite{yang2017fine} & 3D cFPN & \textbf{Ours}  \\
			\midrule[1.1pt]			
			Dice & $0.82\pm0.04$ & $0.84\pm0.04$ & $0.82\pm0.04$ & $0.88\pm0.04$ & \textbf{0.90$\pm$0.03} \\
			\hline
			Jaccard & $0.70\pm0.06$ & $0.73\pm0.06$ & $0.70\pm0.05$ & $0.78\pm0.06$ & \textbf{0.82$\pm$0.04} \\
			\hline
			CC & $0.56\pm0.12$ & $0.63\pm0.11$ & $0.56\pm0.11$ & $0.72\pm0.10$ & \textbf{0.78$\pm$0.08} \\
			\hline
			ADB & $9.58\pm2.65$ & $8.27\pm2.03$ & $5.13\pm1.13$ & $6.12\pm1.88$ & \textbf{3.32$\pm$1.15} \\
			\hline
			$95$HD & $25.11\pm7.83$ & $20.39\pm4.74$ & $11.57\pm2.64$ & $15.11\pm5.03$ & \textbf{8.37$\pm$2.52}  \\
			\hline
			Precision & $0.81\pm0.09$ & $0.83\pm0.08$ & $0.87\pm0.07$ & $0.85\pm0.08$ & \textbf{0.90$\pm$0.06} \\
			\hline
			Recall & $0.85\pm0.09$ & $0.88\pm0.08$ & $0.79\pm0.08$ & \textbf{0.92$\pm$0.06} & $0.91\pm0.04$ \\
			\bottomrule[1.5pt]
		\end{tabular}
	\end{center}
\end{table*}

\begin{table}[t]	
	\caption{\textit{P}-values from Wilcoxon rank-sum tests between our method and other compared methods on different metrics}
	\label{table:t-test}
	\begin{center}
		\begin{tabular}{l|c|c|c|c}
			\toprule[1.5pt]
			Metric & \tabincell{c}{3D FCN\\\textit{vs.}\\Ours} & \tabincell{c}{3D U-Net\\\textit{vs.}\\Ours} & \tabincell{c}{BCRNN\\\textit{vs.}\\Ours} & \tabincell{c}{3D cFPN\\\textit{vs.}\\Ours}\\
			\midrule[1.1pt]			
			Dice & $10^{-12}$ & $10^{-10}$ & $10^{-12}$ & $10^{-3}$ \\
			\hline
			Jaccard & $10^{-12}$ & $10^{-10}$ & $10^{-12}$ & $10^{-3}$ \\
			\hline
			CC & $10^{-12}$ & $10^{-10}$ & $10^{-12}$ & $10^{-3}$ \\
			\hline
			ADB & $10^{-14}$ & $10^{-14}$ & $10^{-8}$ & $10^{-10}$ \\
			\hline
			$95$HD & $10^{-14}$ & $10^{-14}$ & $10^{-7}$ & $10^{-11}$  \\
			\hline
			Precision & $10^{-6}$ & $10^{-6}$ & $0.03$ & $10^{-3}$ \\
			\hline
			Recall & $0.01$ & $0.11$ & $10^{-8}$ & $0.08$ \\
			\bottomrule[1.5pt]
		\end{tabular}
	\end{center}
\end{table}

The metrics employed to quantitatively evaluate segmentation included Dice Similarity Coefficient (Dice), Jaccard Index, Conformity Coefficient (CC), Average Distance of Boundaries (ADB, in voxel), $95\%$ Hausdorff Distance ($95$HD, in voxel), Precision, and Recall \cite{chang2009performance, litjens2014evaluation}.
Metrics of Dice, Jaccard and CC were used to evaluate the similarity between the segmented volume and ground truth \footnote{Dice=2(G$\cap$S)$/$(G+S), Jaccard=(G$\cap$S)$/$(G$\cup$S), CC=2-(G$\cup$S)$/$(G$\cap$S), where S and G denotes the segmented volume and ground truth, respectively.}.
The ADB measured the average over the shortest voxel distances between the segmented volume and ground truth.
The HD is the longest distance over the shortest distances between the segmented volume and ground truth.
Because HD is sensitive to outliers, we used the $95th$ percentile of the asymmetric HD instead of the maximum.
Precision and Recall evaluated segmentations from the aspect of voxel-wise classification accuracy.
All evaluation metrics were calculated in 3D.
A better segmentation shall have smaller ADB and $95$HD, and larger values of all other metrics.

\subsection{Segmentation Performance}

We first qualitatively illustrate the effectiveness of the proposed attention module for the feature refinement.
From Fig.~\ref{fig:attim}, we can observe that our attentive map can indicate how much attention should be paid to the MLF for each SLF, and thus is able to select the useful complementary information from the MLF to refine each SLF correspondingly.

Table~\ref{table:state-of-the-art} summarizes the numerical results of all compared methods.
It can be observed that our method consistently outperforms others on almost all the metrics.
Specifically, our method yielded the mean Dice value of $0.90$, Jaccard of $0.82$, CC of $0.78$, ADB of $3.32$ voxels, $95$HD of $8.37$ voxels, and Precision of $0.90$.
All these results are the best among all compared methods.
Note that our method had the second best mean Recall value among all methods; our customized feature pyramid network achieved the best Recall value.
However, except for the mean Recall value, our attention guided network outperforms the ablation model (i.e., the 3D cFPN) with regard to all the other metrics.
Specifically, as shown in Table~\ref{table:state-of-the-art}, the mean Dice, Jaccard, CC, ADB, $95$HD, and Precision values by the proposed attention guided network are approximately $2.57\%$, $4.58\%$, $8.18\%$, $45.74\%$, $44.61\%$, and $5.85\%$ better than the ablation model without attention modules, respectively.
These comparison results between our method and the 3D cFPN demonstrate that the proposed attention module contributes to the improvement of the TRUS segmentation.
Although our customized 3D FPN architecture already consistently outperforms existing state-of-the-art segmentation methods on most of the metrics by leveraging the useful
multi-level features, the proposed attention module has the capability to more effectively leverage the useful complementary information encoded in the multi-level features to refine themselves for even better segmentation.

\begin{figure} [t]
	\centering
	\includegraphics[width=0.87\linewidth]{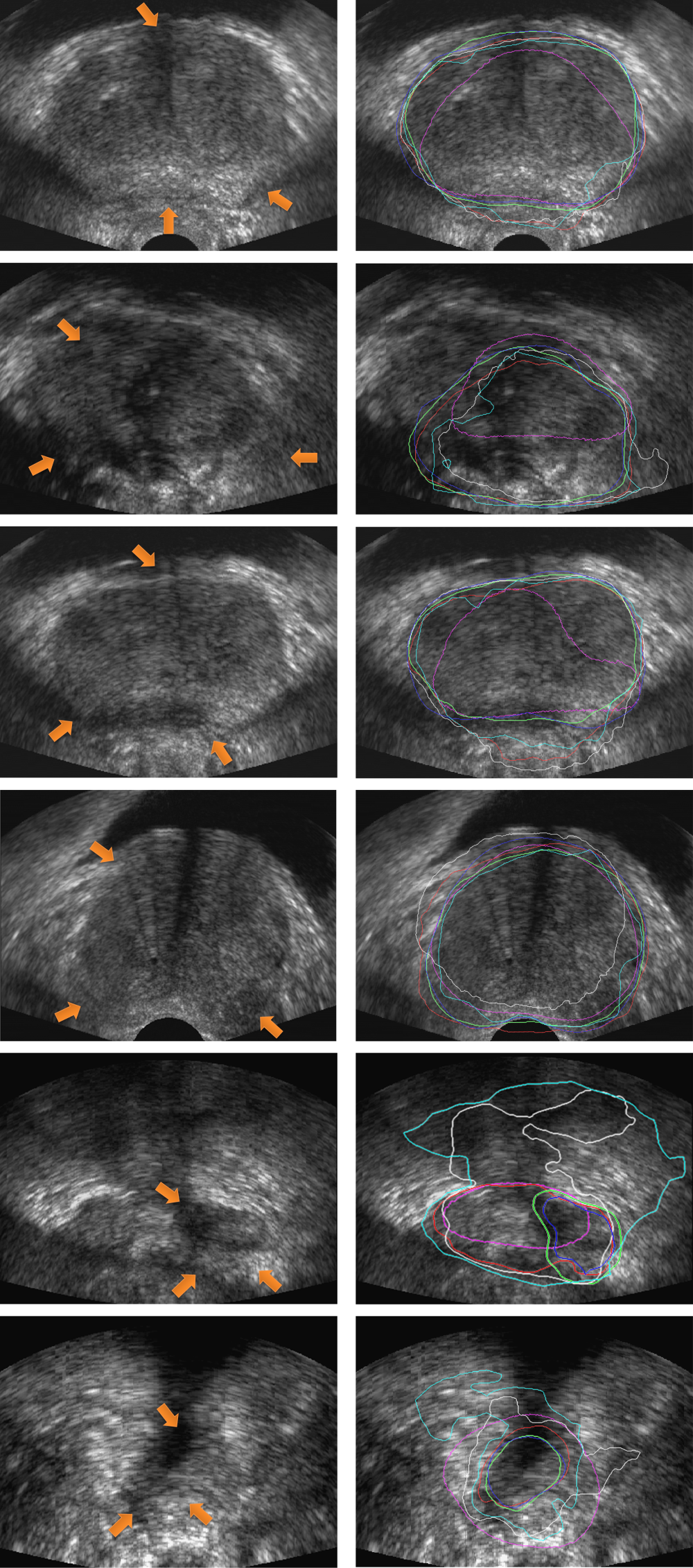}
	\caption{2D visual comparisons of segmented slices from 3D TRUS volumes.
				Left: prostate TRUS slices with orange arrows indicating missing/ambiguous boundaries;
				Right: corresponding segmented prostate boundaries using our method (green), 3D FCN \cite{long2015fully} (cyan), 3D U-Net \cite{ghavami2018automatic} (gray), BCRNN \cite{yang2017fine} (purple) and 3D cFPN (red), respectively.
				Blue contours are ground truths extracted by an experienced clinician.
				Our method has the most similar segmented boundaries to the ground truths.
				Specifically, compared to our ablation study (red contours), the proposed attention module is beneficial to learn more discriminative features indicating real prostate region and boundary.
				(We encourage you to zoom in for better visualization.)
	}
	\label{fig:comparison_real_photos}
\end{figure}

To investigate the statistical significance of the proposed method over compared methods on each of the metrics, a series of statistical analyses are conducted.
First, the one-way analysis of variance (ANOVA) \cite{Neter1985Applied} is performed to evaluate if the metric results of different methods are statistically different.
The resulting $F_{Dice} = 34.85$, $F_{Jaccard} = 36.71$, $F_{CC} = 32.22$, $F_{ADB} = 71.73$, $F_{95HD} = 73.83$, $F_{Precision} = 7.88$, and $F_{Recall} = 18.80$, respectively;
all are larger than the same $F_{critical} (= 2.42)$, indicating that the differences between each of the metrics from the five methods are statistically significant.
Based on the observations from ANOVA, the Wilcoxon rank-sum test is further employed to compare the segmentation performances between our method and other compared methods.
Table~\ref{table:t-test} lists the $p$-values from Wilcoxon rank-sum tests between our method and other compared methods on different metrics.
By observing Table~\ref{table:t-test}, it can be concluded that the null hypotheses for the four comparing pairs on the metrics of Dice, Jaccard, CC, ADB, $95$HD, and Precision are not accepted at the $0.05$ level.
As a result, our method can be regarded as significantly better than the other four compared methods on these evaluation metrics.
It is worth noting that the $p$-values of 3D U-Net-Ours and 3D cFPN-Ours on metric Recall are beyond the $0.05$ level, which indicates that our method, 3D U-Net and 3D cFPN achieve similar performance with regard to the Recall evaluation.
In general, the results shown in Tables~\ref{table:state-of-the-art} and~\ref{table:t-test} prove the effect of our attention guided network on the accurate TRUS segmentation.

\begin{figure*} [t]
	\centering
	\includegraphics[width=0.99\linewidth]{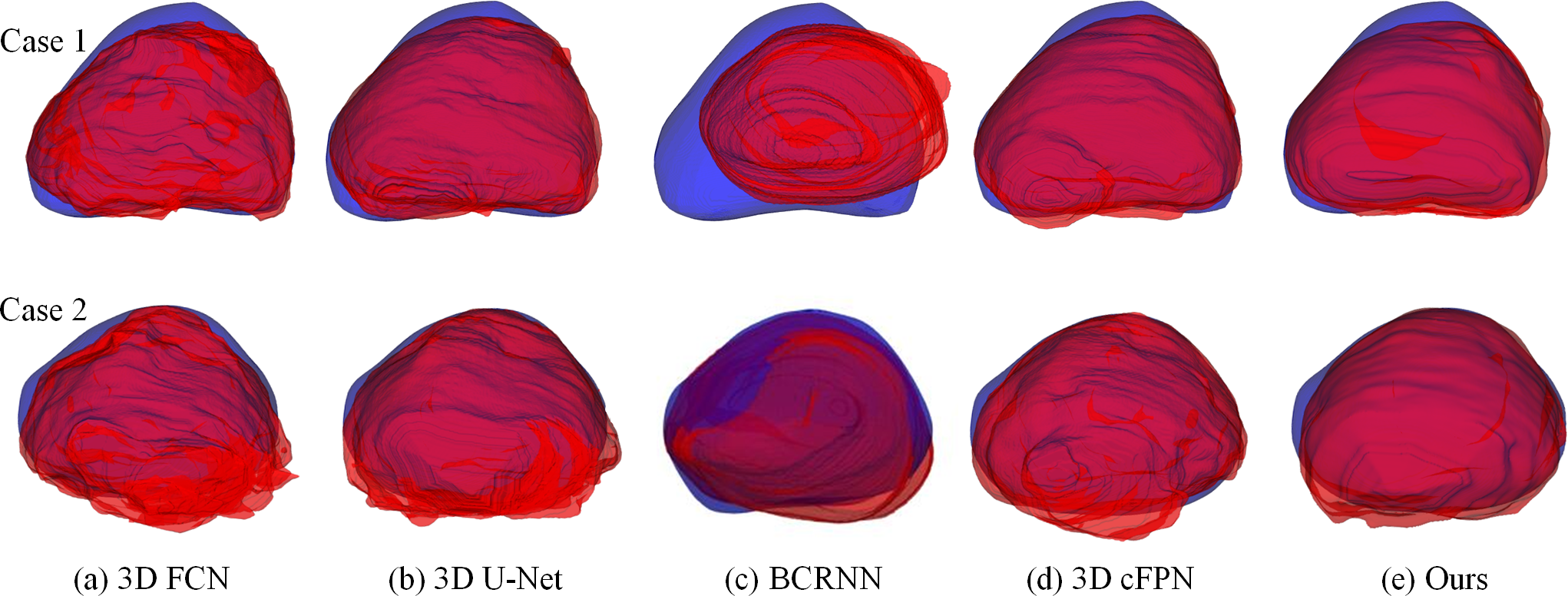}
	\caption{3D visualization of the segmentation results on two TRUS volumes. Rows indicate segmentation results on different TRUS data. Columns indicate the comparisons between ground truth (blue surface) and segmented surfaces (red) using (a) 3D FCN \cite{long2015fully}, (b) 3D U-Net \cite{ghavami2018automatic}, (c) BCRNN \cite{yang2017fine}, (d) 3D cFPN, and (e) our method, respectively. Our method has the most similar segmented surfaces to the ground truths.}
	\label{fig:3doverlap}
\end{figure*}

\begin{figure*} [t]
	\centering
	\includegraphics[width=0.99\linewidth]{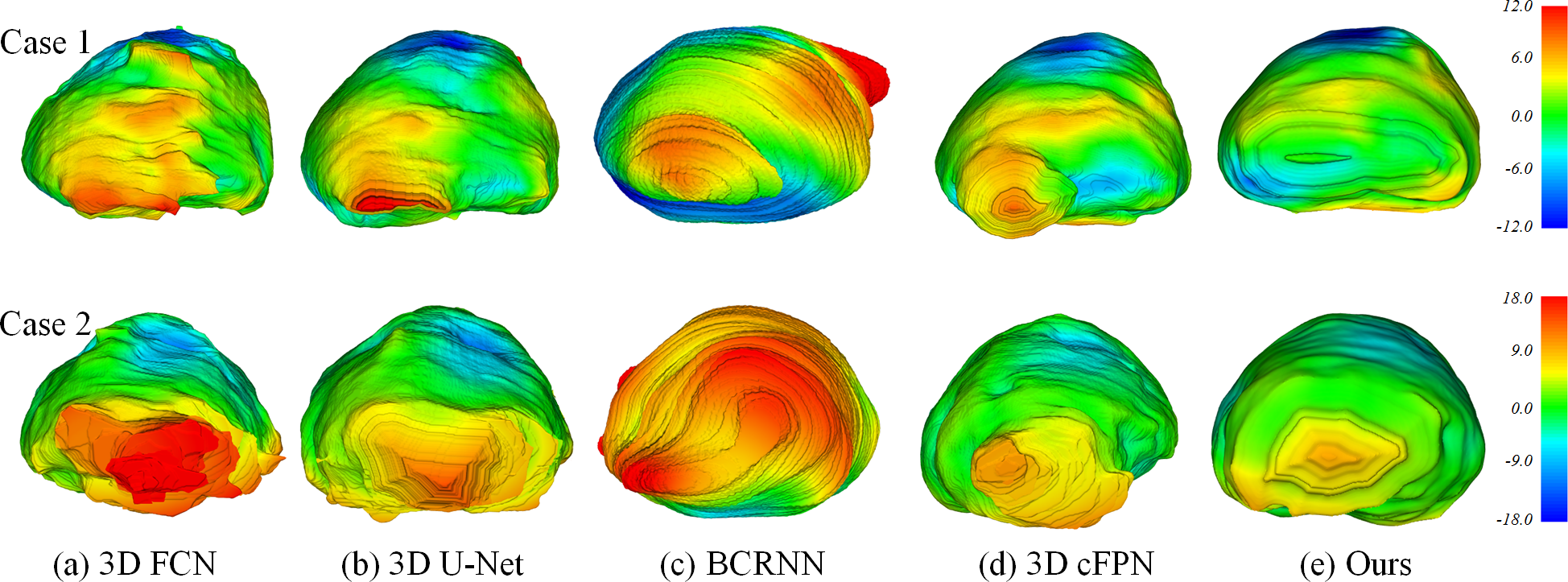}
	\caption{3D visualization of the surface distance (in voxel) between segmented surface and ground truth. Different colors represent different surface distances. Rows indicate segmented surfaces on different TRUS data. Columns indicate the segmented surfaces obtained by (a) 3D FCN \cite{long2015fully}, (b) 3D U-Net \cite{ghavami2018automatic}, (c) BCRNN \cite{yang2017fine}, (d) 3D cFPN, and (e) our method, respectively. Our method consistently performs well on the whole prostate surface.
	}
	\label{fig:3ddistance}
\end{figure*}

Figs.~\ref{fig:comparison_real_photos},~\ref{fig:3doverlap} and~\ref{fig:3ddistance} visualize some segmentation results in 2D and 3D, respectively.
Fig.~\ref{fig:comparison_real_photos} compares some segmented boundaries by different methods in 2D TRUS slices.
Apparently, our method obtains the most similar segmented boundaries (green contours) to the ground truths (blue contours).
Furthermore, as shown in Fig.~\ref{fig:comparison_real_photos}, our method can successfully infer the missing/ambiguous boundaries, whereas other compared methods including 3D cFPN tend to fail in generating high-quality segmentations when there are ambiguous/missing boundaries in TRUS images.
These comparisons demonstrate that the proposed deep attentive features can efficiently aggregate complementary multi-level information for accurate representation of the prostate tissues.
Figs.~\ref{fig:3doverlap} and~\ref{fig:3ddistance} visualize 3D segmentation results by different methods on two TRUS volumes.
As shown in Fig.~\ref{fig:3doverlap}, our method has the most similar segmented surfaces to the ground truths (blue surfaces).
Fig.~\ref{fig:3ddistance} further depicts the corresponding surface distance between segmented surfaces and ground truths.
It can be observed that our method consistently achieves accurate and robust segmentation covering the whole prostate region.

Given the $170 \times 132 \times 80$ voxels input 3D TRUS volume, the average computational times needed to perform a whole prostate segmentation for 3D FCN, 3D U-Net, BCRNN, 3D cFPN and our method are $1.10$, $0.34$, $31.09$, $0.24$ and $0.30$ seconds, respectively. Our method is faster than the 3D FCN, 3D U-Net and BCRNN.

\section{Discussion}
\label{sec:Discussions}

In this paper, an attention guided neural network which generates attentive features for the segmentation of 3D TRUS volumes is presented.
Accurate and robust prostate segmentation in TRUS images remains very challenging mainly due to the missing/ambiguous boundary of the prostate in TRUS.
Conventional methods mainly employ prior shape information to constrain the segmentation, or design hand-crafted features to identify prostate regions, which generally tend to fail in faithfully delineating boundaries when there are missing/ambiguous boundaries in TRUS images \cite{ghose2012survey}.
Recently, since convolutional neural network approaches have demonstrated to be very powerful to learn multi-level features for the effective object segmentation \cite{chen2018deeplab}, we are motivated to develop a CNN based method to tackle the challenging issues in TRUS segmentation.
To the best of our knowledge, we are the pioneer to utilize 3D CNN with attention mechanisms to refine multi-level features for the better TRUS segmentation.

Deep convolutional neural networks have achieved superior performance in many image computing and vision tasks, due to the advantage of generating multi-level features consisting of abundant semantic and fine information.
However, how to leverage the complementary advantages of multi-level features and to learn more discriminative features for image segmentation remains the key issue to be addressed.
As shown in Figs.~\ref{fig:mlf_daf},~\ref{fig:attim} and~\ref{fig:comparison_real_photos}, directly applying multi-level convolutional features without distinction for TRUS segmentation tends to include non-prostate regions (due to low-level details from shallow layers) or lose details of prostate boundaries (due to high-level semantics from deep layers).
In order to address this issue, we propose an attention guided network to select more discriminative features for TRUS segmentation.
Our attention module leverages the MLF as a fruitful feature pool to refine each SLF, by learning a set of weights to indicate the importance of MLF for specific SLF.
Table~\ref{table:state-of-the-art} and Figs.~\ref{fig:comparison_real_photos},~\ref{fig:3doverlap} and~\ref{fig:3ddistance} all demonstrate that our attention module is useful to improve multi-level features for 3D TRUS segmentation.
More generally, the proposed attention module provides a feasible solution to effectively incorporate details at low levels and semantics at high levels for better feature representation.
Thus as a generic feature refinement architecture, our attention module is potentially useful to become a beneficial component in other segmentation/detection networks for their performance improvement.

Considering the issue of missing/ambiguous prostate boundary in TRUS images, we adopt a hybrid loss function, which combines binary cross-entropy loss and Dice loss for our segmentation network.
The binary cross-entropy loss is preferred in preserving boundary details while the Dice loss emphasizes global shape similarity to generate compact segmentation.
Therefore the hybrid loss is beneficial to leverage both local and global shape similarity.
This hybrid loss could be useful for other segmentation tasks and we will further explore it in our future work.

Although the proposed method achieves satisfactory performance in the experiments, there is still one important limitation in this study.
The experiments were based on a four-fold cross-validation study with only forty TRUS volumes.
In each fold, test data were held out while the data from the remaining patients were used in training.
The cross-validation was also to identify hyper-parameters that generalize well across the samples we learn from in each fold.
Such cross-validation approach on forty samples may have caused over-fitting to training samples.
As a result, future studies will focus on evaluating the generalizability of the approach on a larger dataset by properly dividing data to mutually exclusive training, validation and test subsets.

\section{Conclusion}
\label{sec:Conclusion}

This paper develops a 3D attention guided neural network with a novel scheme for prostate segmentation in 3D transrectal ultrasound images by harnessing the deep attentive features.
Our key idea is to select the useful complementary information from the multi-level features to refine the features at each individual layer.
We achieve this by developing an attention module, which can automatically learn a set of weights to indicate the importance of the features in MLF for each individual layer by using an attention mechanism.
To the best of our knowledge, we are the first to utilize attention mechanisms to refine multi-level features for the better 3D TRUS segmentation.
Experiments on challenging TRUS volumes show that our segmentation using deep attentive features achieves satisfactory performance.
In addition, the proposed attention mechanism is a general strategy to aggregate multi-level features and has the potential to be used for other medical image segmentation and detection tasks.

\section*{Acknowledgment}
The authors would like to thank the Associate Editor and the anonymous reviewers for their constructive comments.

\ifCLASSOPTIONcaptionsoff
  \newpage
\fi



\bibliographystyle{IEEEtran}
\bibliography{TMI-2019-0221R2}

\begin{thebibliography}{10}
\providecommand{\url}[1]{#1}
\csname url@samestyle\endcsname
\providecommand{\newblock}{\relax}
\providecommand{\bibinfo}[2]{#2}
\providecommand{\BIBentrySTDinterwordspacing}{\spaceskip=0pt\relax}
\providecommand{\BIBentryALTinterwordstretchfactor}{4}
\providecommand{\BIBentryALTinterwordspacing}{\spaceskip=\fontdimen2\font plus
\BIBentryALTinterwordstretchfactor\fontdimen3\font minus
  \fontdimen4\font\relax}
\providecommand{\BIBforeignlanguage}[2]{{%
\expandafter\ifx\csname l@#1\endcsname\relax
\typeout{** WARNING: IEEEtran.bst: No hyphenation pattern has been}%
\typeout{** loaded for the language `#1'. Using the pattern for}%
\typeout{** the default language instead.}%
\else
\language=\csname l@#1\endcsname
\fi
#2}}
\providecommand{\BIBdecl}{\relax}
\BIBdecl

\bibitem{siegel2018cancer}
R.~L. Siegel, K.~D. Miller, and A.~Jemal, ``Cancer statistics, 2018,''
  \emph{CA: A Cancer Journal for Clinicians}, vol.~68, no.~1, pp. 7--30, 2018.

\bibitem{pinto2011imaging}
F.~Pinto, A.~Totaro, A.~Calarco, E.~Sacco, A.~Volpe, M.~Racioppi,
  A.~D’Addessi, G.~Gulino, and P.~Bassi, ``Imaging in prostate cancer
  diagnosis: present role and future perspectives,'' \emph{Urologia
  Internationalis}, vol.~86, no.~4, pp. 373--382, 2011.

\bibitem{hricak2007imaging}
H.~Hricak, P.~L. Choyke, S.~C. Eberhardt, S.~A. Leibel, and P.~T. Scardino,
  ``Imaging prostate cancer: a multidisciplinary perspective,''
  \emph{Radiology}, vol. 243, no.~1, pp. 28--53, 2007.

\bibitem{wang2016towards}
Y.~Wang, J.-Z. Cheng, D.~Ni, M.~Lin, J.~Qin, X.~Luo, M.~Xu, X.~Xie, and P.~A.
  Heng, ``Towards personalized statistical deformable model and hybrid point
  matching for robust {MR}-{TRUS} registration,'' \emph{IEEE Transactions on
  Medical Imaging}, vol.~35, no.~2, pp. 589--604, 2016.

\bibitem{yan2010discrete}
P.~Yan, S.~Xu, B.~Turkbey, and J.~Kruecker, ``Discrete deformable model guided
  by partial active shape model for {TRUS} image segmentation,'' \emph{IEEE
  Transactions on Biomedical Engineering}, vol.~57, no.~5, pp. 1158--1166,
  2010.

\bibitem{davis2012american}
B.~J. Davis, E.~M. Horwitz, W.~R. Lee, J.~M. Crook, R.~G. Stock, G.~S. Merrick,
  W.~M. Butler, P.~D. Grimm, N.~N. Stone, L.~Potters \emph{et~al.}, ``American
  brachytherapy society consensus guidelines for transrectal ultrasound-guided
  permanent prostate brachytherapy,'' \emph{Brachytherapy}, vol.~11, no.~1, pp.
  6--19, 2012.

\bibitem{bahn2002targeted}
D.~K. Bahn, F.~Lee, R.~Badalament, A.~Kumar, J.~Greski, and M.~Chernick,
  ``Targeted cryoablation of the prostate: 7-year outcomes in the primary
  treatment of prostate cancer,'' \emph{Urology}, vol.~60, no.~2, pp. 3--11,
  2002.

\bibitem{hu2012mr}
Y.~Hu, H.~U. Ahmed, Z.~Taylor, C.~Allen, M.~Emberton, D.~Hawkes, and
  D.~Barratt, ``{MR} to ultrasound registration for image-guided prostate
  interventions,'' \emph{Medical Image Analysis}, vol.~16, no.~3, pp. 687--703,
  2012.

\bibitem{wang2018online}
Y.~Wang, Q.~Zheng, and P.~A. Heng, ``Online robust projective dictionary
  learning: Shape modeling for {MR}-{TRUS} registration,'' \emph{IEEE
  Transactions on Medical Imaging}, vol.~37, no.~4, pp. 1067--1078, 2018.

\bibitem{noble2006ultrasound}
J.~A. Noble and D.~Boukerroui, ``Ultrasound image segmentation: a survey,''
  \emph{IEEE Transactions on Medical Imaging}, vol.~25, no.~8, pp. 987--1010,
  2006.

\bibitem{ghose2012survey}
S.~Ghose, A.~Oliver, R.~Mart{\'\i}, X.~Llad{\'o}, J.~C. Vilanova, J.~Freixenet,
  J.~Mitra, D.~Sidib{\'e}, and F.~Meriaudeau, ``A survey of prostate
  segmentation methodologies in ultrasound, magnetic resonance and computed
  tomography images,'' \emph{Computer Methods and Programs in Biomedicine},
  vol. 108, no.~1, pp. 262--287, 2012.

\bibitem{ladak2000prostate}
H.~M. Ladak, F.~Mao, Y.~Wang, D.~B. Downey, D.~A. Steinman, and A.~Fenster,
  ``Prostate boundary segmentation from 2{D} ultrasound images,'' \emph{Medical
  Physics}, vol.~27, no.~8, pp. 1777--1788, 2000.

\bibitem{pathak2000edge}
S.~D. Pathak, D.~Haynor, and Y.~Kim, ``Edge-guided boundary delineation in
  prostate ultrasound images,'' \emph{IEEE Transactions on Medical Imaging},
  vol.~19, no.~12, pp. 1211--1219, 2000.

\bibitem{ghanei2001three}
A.~Ghanei, H.~Soltanian-Zadeh, A.~Ratkewicz, and F.-F. Yin, ``A
  three-dimensional deformable model for segmentation of human prostate from
  ultrasound images,'' \emph{Medical Physics}, vol.~28, no.~10, pp. 2147--2153,
  2001.

\bibitem{shen2003segmentation}
D.~Shen, Y.~Zhan, and C.~Davatzikos, ``Segmentation of prostate boundaries from
  ultrasound images using statistical shape model,'' \emph{IEEE Transactions on
  Medical Imaging}, vol.~22, no.~4, pp. 539--551, 2003.

\bibitem{wang2003semiautomatic}
Y.~Wang, H.~N. Cardinal, D.~B. Downey, and A.~Fenster, ``Semiautomatic
  three-dimensional segmentation of the prostate using two-dimensional
  ultrasound images,'' \emph{Medical Physics}, vol.~30, no.~5, pp. 887--897,
  2003.

\bibitem{hu2003prostate}
N.~Hu, D.~B. Downey, A.~Fenster, and H.~M. Ladak, ``Prostate boundary
  segmentation from 3{D} ultrasound images,'' \emph{Medical Physics}, vol.~30,
  no.~7, pp. 1648--1659, 2003.

\bibitem{gong2004parametric}
L.~Gong, S.~D. Pathak, D.~R. Haynor, P.~S. Cho, and Y.~Kim, ``Parametric shape
  modeling using deformable superellipses for prostate segmentation,''
  \emph{IEEE Transactions on Medical Imaging}, vol.~23, no.~3, pp. 340--349,
  2004.

\bibitem{badiei2006prostate}
S.~Badiei, S.~E. Salcudean, J.~Varah, and W.~J. Morris, ``Prostate segmentation
  in 2{D} ultrasound images using image warping and ellipse fitting,'' in
  \emph{International Conference on Medical Image Computing and
  Computer-Assisted Intervention}.\hskip 1em plus 0.5em minus 0.4em\relax
  Springer, 2006, pp. 17--24.

\bibitem{tutar2006semiautomatic}
I.~B. Tutar, S.~D. Pathak, L.~Gong, P.~S. Cho, K.~Wallner, and Y.~Kim,
  ``Semiautomatic 3-{D} prostate segmentation from {TRUS} images using
  spherical harmonics,'' \emph{IEEE Transactions on Medical Imaging}, vol.~25,
  no.~12, pp. 1645--1654, 2006.

\bibitem{zhan2006deformable}
Y.~Zhan and D.~Shen, ``Deformable segmentation of 3-{D} ultrasound prostate
  images using statistical texture matching method,'' \emph{IEEE Transactions
  on Medical Imaging}, vol.~25, no.~3, pp. 256--272, 2006.

\bibitem{yann2011adaptively}
P.~Yan, S.~Xu, B.~Turkbey, and J.~Kruecker, ``Adaptively learning local shape
  statistics for prostate segmentation in ultrasound,'' \emph{IEEE Transactions
  on Biomedical Engineering}, vol.~58, no.~3, pp. 633--641, 2011.

\bibitem{ghose2013supervised}
S.~Ghose, A.~Oliver, J.~Mitra, R.~Mart{\'\i}, X.~Llad{\'o}, J.~Freixenet,
  D.~Sidib{\'e}, J.~C. Vilanova, J.~Comet, and F.~Meriaudeau, ``A supervised
  learning framework of statistical shape and probability priors for automatic
  prostate segmentation in ultrasound images,'' \emph{Medical Image Analysis},
  vol.~17, no.~6, pp. 587--600, 2013.

\bibitem{qiu2014prostate}
W.~Qiu, J.~Yuan, E.~Ukwatta, Y.~Sun, M.~Rajchl, and A.~Fenster, ``Prostate
  segmentation: an efficient convex optimization approach with axial symmetry
  using 3-{D} {TRUS} and {MR} images,'' \emph{IEEE Transactions on Medical
  Imaging}, vol.~33, no.~4, pp. 947--960, 2014.

\bibitem{santiago20152d}
C.~Santiago, J.~C. Nascimento, and J.~S. Marques, ``2{D} segmentation using a
  robust active shape model with the {EM} algorithm,'' \emph{IEEE Transactions
  on Image Processing}, vol.~24, no.~8, pp. 2592--2601, 2015.

\bibitem{wu2015robust}
P.~Wu, Y.~Liu, Y.~Li, and B.~Liu, ``Robust prostate segmentation using
  intrinsic properties of {TRUS} images,'' \emph{IEEE Transactions on Medical
  Imaging}, vol.~34, no.~6, pp. 1321--1335, 2015.

\bibitem{li2016segmentation}
X.~Li, C.~Li, A.~Fedorov, T.~Kapur, and X.~Yang, ``Segmentation of prostate
  from ultrasound images using level sets on active band and intensity
  variation across edges,'' \emph{Medical Physics}, vol.~43, no. 6Part1, pp.
  3090--3103, 2016.

\bibitem{yang20163d}
X.~Yang, P.~J. Rossi, A.~B. Jani, H.~Mao, W.~J. Curran, and T.~Liu, ``3{D}
  transrectal ultrasound ({TRUS}) prostate segmentation based on optimal
  feature learning framework,'' in \emph{Medical Imaging 2016: Image
  Processing}, vol. 9784.\hskip 1em plus 0.5em minus 0.4em\relax International
  Society for Optics and Photonics, 2016, p. 97842F.

\bibitem{zhu2017non}
L.~Zhu, C.-W. Fu, M.~S. Brown, and P.-A. Heng, ``A non-local low-rank framework
  for ultrasound speckle reduction,'' in \emph{Proceedings of the IEEE
  Conference on Computer Vision and Pattern Recognition}, 2017, pp. 5650--5658.

\bibitem{ma2017random}
L.~Ma, R.~Guo, Z.~Tian, and B.~Fei, ``A random walk-based segmentation
  framework for 3{D} ultrasound images of the prostate,'' \emph{Medical
  Physics}, vol.~44, no.~10, pp. 5128--5142, 2017.

\bibitem{ciresan2012deep}
D.~Ciresan, A.~Giusti, L.~M. Gambardella, and J.~Schmidhuber, ``Deep neural
  networks segment neuronal membranes in electron microscopy images,'' in
  \emph{Advances in Neural Information Processing Systems}, 2012, pp.
  2843--2851.

\bibitem{schmidhuber2015deep}
J.~Schmidhuber, ``Deep learning in neural networks: An overview,'' \emph{Neural
  Networks}, vol.~61, pp. 85--117, 2015.

\bibitem{long2015fully}
J.~Long, E.~Shelhamer, and T.~Darrell, ``Fully convolutional networks for
  semantic segmentation,'' in \emph{Proceedings of the IEEE Conference on
  Computer Vision and Pattern Recognition}, 2015, pp. 3431--3440.

\bibitem{ronneberger2015u}
O.~Ronneberger, P.~Fischer, and T.~Brox, ``U-net: Convolutional networks for
  biomedical image segmentation,'' in \emph{International Conference on Medical
  Image Computing and Computer-Assisted Intervention}.\hskip 1em plus 0.5em
  minus 0.4em\relax Springer, 2015, pp. 234--241.

\bibitem{liskowski2016segmenting}
P.~Liskowski and K.~Krawiec, ``Segmenting retinal blood vessels with deep
  neural networks,'' \emph{IEEE Transactions on Medical Imaging}, vol.~35,
  no.~11, pp. 2369--2380, 2016.

\bibitem{havaei2017brain}
M.~Havaei, A.~Davy, D.~Warde-Farley, A.~Biard, A.~Courville, Y.~Bengio, C.~Pal,
  P.-M. Jodoin, and H.~Larochelle, ``Brain tumor segmentation with deep neural
  networks,'' \emph{Medical Image Analysis}, vol.~35, pp. 18--31, 2017.

\bibitem{chen2018deeplab}
L.-C. Chen, G.~Papandreou, I.~Kokkinos, K.~Murphy, and A.~L. Yuille, ``Deeplab:
  Semantic image segmentation with deep convolutional nets, atrous convolution,
  and fully connected {CRF}s,'' \emph{IEEE Transactions on Pattern Analysis and
  Machine Intelligence}, vol.~40, no.~4, pp. 834--848, 2018.

\bibitem{7353170}
Y.~Guo, Y.~Gao, and D.~Shen, ``Deformable {MR} prostate segmentation via deep
  feature learning and sparse patch matching,'' \emph{IEEE Transactions on
  Medical Imaging}, vol.~35, no.~4, pp. 1077--1089, 2016.

\bibitem{ghavami2018automatic}
N.~Ghavami, Y.~Hu, E.~Bonmati, R.~Rodell, E.~Gibson, C.~Moore, and D.~Barratt,
  ``Automatic slice segmentation of intraoperative transrectal ultrasound
  images using convolutional neural networks,'' in \emph{Medical Imaging 2018:
  Image-Guided Procedures, Robotic Interventions, and Modeling}, vol.
  10576.\hskip 1em plus 0.5em minus 0.4em\relax International Society for
  Optics and Photonics, 2018, p. 1057603.

\bibitem{ghavami2018integration}
N.~Ghavami, Y.~Hu, E.~Bonmati, R.~Rodell, E.~Gibson, and et~al, ``Integration
  of spatial information in convolutional neural networks for automatic
  segmentation of intraoperative transrectal ultrasound images,'' \emph{Journal
  of Medical Imaging}, vol.~6, no.~1, p. 011003, 2018.

\bibitem{yang2017fine}
X.~Yang, L.~Yu, L.~Wu, Y.~Wang, D.~Ni, J.~Qin, and P.-A. Heng, ``Fine-grained
  recurrent neural networks for automatic prostate segmentation in ultrasound
  images,'' in \emph{AAAI Conference on Artificial Intelligence}, 2017, pp.
  1633--1639.

\bibitem{karimi2018accurate}
D.~Karimi, Q.~Zeng, P.~Mathur, A.~Avinash, S.~Mahdavi, I.~Spadinger,
  P.~Abolmaesumi, and S.~Salcudean, ``Accurate and robust segmentation of the
  clinical target volume for prostate brachytherapy,'' in \emph{International
  Conference on Medical Image Computing and Computer-Assisted
  Intervention}.\hskip 1em plus 0.5em minus 0.4em\relax Springer, 2018, pp.
  531--539.

\bibitem{anas2017clinical}
E.~M.~A. Anas, S.~Nouranian, S.~S. Mahdavi, I.~Spadinger, W.~J. Morris, S.~E.
  Salcudean, P.~Mousavi, and P.~Abolmaesumi, ``Clinical target-volume
  delineation in prostate brachytherapy using residual neural networks,'' in
  \emph{International Conference on Medical Image Computing and
  Computer-Assisted Intervention}.\hskip 1em plus 0.5em minus 0.4em\relax
  Springer, 2017, pp. 365--373.

\bibitem{anas2018deep}
E.~M.~A. Anas, P.~Mousavi, and P.~Abolmaesumi, ``A deep learning approach for
  real time prostate segmentation in freehand ultrasound guided biopsy,''
  \emph{Medical Image Analysis}, vol.~48, pp. 107--116, 2018.

\bibitem{wang2018deep}
Y.~Wang, Z.~Deng, X.~Hu, L.~Zhu, X.~Yang, X.~Xu, P.-A. Heng, and D.~Ni, ``Deep
  attentional features for prostate segmentation in ultrasound,'' in
  \emph{International Conference on Medical Image Computing and
  Computer-Assisted Intervention}.\hskip 1em plus 0.5em minus 0.4em\relax
  Springer, 2018, pp. 523--530.

\bibitem{xie2017aggregated}
S.~Xie, R.~Girshick, P.~Doll{\'a}r, Z.~Tu, and K.~He, ``Aggregated residual
  transformations for deep neural networks,'' in \emph{Proceedings of the IEEE
  International on Computer Vision and Pattern Recognition}.\hskip 1em plus
  0.5em minus 0.4em\relax IEEE, 2017, pp. 5987--5995.

\bibitem{yu2015multi}
F.~Yu and V.~Koltun, ``Multi-scale context aggregation by dilated
  convolutions,'' \emph{arXiv preprint arXiv:1511.07122}, 2015.

\bibitem{lin2017feature}
T.-Y. Lin, P.~Doll{\'a}r, R.~B. Girshick, K.~He, B.~Hariharan, and S.~J.
  Belongie, ``Feature pyramid networks for object detection,'' in
  \emph{Proceedings of the IEEE International on Computer Vision and Pattern
  Recognition}, 2017, pp. 2117--2125.

\bibitem{xie2015holistically}
S.~Xie and Z.~Tu, ``Holistically-nested edge detection,'' in \emph{Proceedings
  of the IEEE International Conference on Computer Vision}, 2015, pp.
  1395--1403.

\bibitem{dou20173d}
Q.~Dou, L.~Yu, H.~Chen, Y.~Jin, X.~Yang, J.~Qin, and P.-A. Heng, ``3{D} deeply
  supervised network for automated segmentation of volumetric medical images,''
  \emph{Medical Image Analysis}, vol.~41, pp. 40--54, 2017.

\bibitem{chen2017rethinking}
L.-C. Chen, G.~Papandreou, F.~Schroff, and H.~Adam, ``Rethinking atrous
  convolution for semantic image segmentation,'' \emph{arXiv preprint
  arXiv:1706.05587}, 2017.

\bibitem{Wu_2018_ECCV}
Y.~Wu and K.~He, ``Group normalization,'' in \emph{Proceedings of The European
  Conference on Computer Vision}, September 2018.

\bibitem{hu2018squeeze}
J.~Hu, L.~Shen, and G.~Sun, ``Squeeze-and-excitation networks,'' in
  \emph{Proceedings of the IEEE Conference on Computer Vision and Pattern
  Recognition}, 2018, pp. 7132--7141.

\bibitem{zhao2018psanet}
H.~Zhao, Y.~Zhang, S.~Liu, J.~Shi, C.~Change~Loy, D.~Lin, and J.~Jia,
  ``{PSAN}et: Point-wise spatial attention network for scene parsing,'' in
  \emph{Proceedings of the European Conference on Computer Vision (ECCV)},
  2018, pp. 267--283.

\bibitem{he2015delving}
K.~He, X.~Zhang, S.~Ren, and J.~Sun, ``Delving deep into rectifiers: Surpassing
  human-level performance on imagenet classification,'' in \emph{Proceedings of
  the IEEE International Conference on Computer Vision}, 2015, pp. 1026--1034.

\bibitem{milletari2016v}
F.~Milletari, N.~Navab, and S.-A. Ahmadi, ``V-net: Fully convolutional neural
  networks for volumetric medical image segmentation,'' in \emph{2016 Fourth
  International Conference on 3D Vision (3DV)}.\hskip 1em plus 0.5em minus
  0.4em\relax IEEE, 2016, pp. 565--571.

\bibitem{Kingma2014Adam}
D.~P. Kingma and J.~Ba, ``Adam: A method for stochastic optimization,''
  \emph{Computer Science}, 2014.

\bibitem{chang2009performance}
H.-H. Chang, A.~H. Zhuang, D.~J. Valentino, and W.-C. Chu, ``Performance
  measure characterization for evaluating neuroimage segmentation algorithms,''
  \emph{Neuroimage}, vol.~47, no.~1, pp. 122--135, 2009.

\bibitem{litjens2014evaluation}
G.~Litjens, R.~Toth, W.~van~de Ven, C.~Hoeks, S.~Kerkstra, B.~van Ginneken,
  G.~Vincent, G.~Guillard, N.~Birbeck, J.~Zhang \emph{et~al.}, ``Evaluation of
  prostate segmentation algorithms for {MRI}: the {PROMISE}12 challenge,''
  \emph{Medical Image Analysis}, vol.~18, no.~2, pp. 359--373, 2014.

\bibitem{Neter1985Applied}
J.~Neter, W.~Wasserman, and M.~H. Kutner, \emph{Applied linear statistical
  models: regression, analysis of variance, and experimental designs; third
  edition}.\hskip 1em plus 0.5em minus 0.4em\relax R.D. Irwin, 1985.

\end{thebibliography}
%

%








\end{document}